\documentclass[aps,prd,10pt,twocolumn,superscriptaddress,floatfix,nofootinbib,amsmath,amssymb,altaffilletter,preprintnumbers]{revtex4-1}

\usepackage{graphicx}
\usepackage{dcolumn}
\usepackage{bm}
\usepackage{tensor}
\usepackage{slashed}
\usepackage{multirow}
\usepackage{soul}
\usepackage{amsmath}
\usepackage{mathrsfs}
\usepackage{amssymb}
\usepackage{yfonts}
\usepackage{color}
\usepackage{xspace}
\usepackage{url}
\usepackage{verbatim}
\usepackage{mathtools}
\usepackage{upgreek}
\usepackage{amstext}
\usepackage{booktabs}
\usepackage{tabulary}
\usepackage{tabularx}
\usepackage{etoolbox}
\usepackage[utf8]{inputenc}
\usepackage[dvipsnames,table]{xcolor}
\newcolumntype{Y}{>{\centering\arraybackslash}X}
\usepackage[colorlinks=true,pagebackref=true,pdfstartview=FitV,breaklinks=true]{hyperref}
\hypersetup{urlcolor=BlueViolet,
	    citecolor=Plum,
	    linkcolor=PineGreen}
\usepackage[official]{eurosym}
\definecolor{lightgray}{rgb}{0.9,0.9,0.9}	    
\definecolor{green}{rgb}{0,0.5,0}
\definecolor{red}{rgb}{1,0,0}
\definecolor{blue}{rgb}{0,0,0.5}

\begin{document}

\title{The high-velocity dark matter halo of the Milky Way in light of the LZ 248~keV event}

\author{Ciaran A.~J.~O'Hare}
\affiliation{University of Sydney}

\begin{abstract}
The interpretation of high-energy nuclear recoil events in direct-detection experiments can depend sensitively on the poorly understood high-speed tail of the Galactic dark matter velocity distribution. This fact has been brought into sharper focus recently by an anomalous nuclear-recoil event at $E_R\approx248~\mathrm{keV}$ observed by the LZ experiment. If interpreted as caused by a dark-matter-induced recoil, kinematic constraints on many models imply this particle would have to emerge from the high-speed tail. Here, we re-assess this astrophysical uncertainty using Milky Way analogues from the IllustrisTNG and FIRE simulations. We find that the maximum laboratory-frame DM speed on June 16, when the event was observed, is $792.7^{+82.9}_{-95.6}\, \mathrm{km/s}$ (median and 95\% containment across simulations), with a median very close to the fiducial SHM value. Modelling the tail of the velocity distribution in three dimensions is complicated by the fact that these simulated dark matter halos are anisotropic and are generically seen to spin in the same direction as the baryonic disk. However, we find that this does not reduce the maximum laboratory-frame DM speed because the approximately Gaussian azimuthal velocity distribution is negatively \textit{skewed} by this co-rotation, rather than shifted by it. We illustrate these results in the context of an inelastic dark matter interpretation of the LZ event.
\end{abstract}

\maketitle

\section{Introduction}
\label{sec:introduction}

\begin{figure*}[t]
    \centering
    \includegraphics[height=0.42\linewidth,trim={0cm 0.0cm 2cm 0.0cm}, clip]{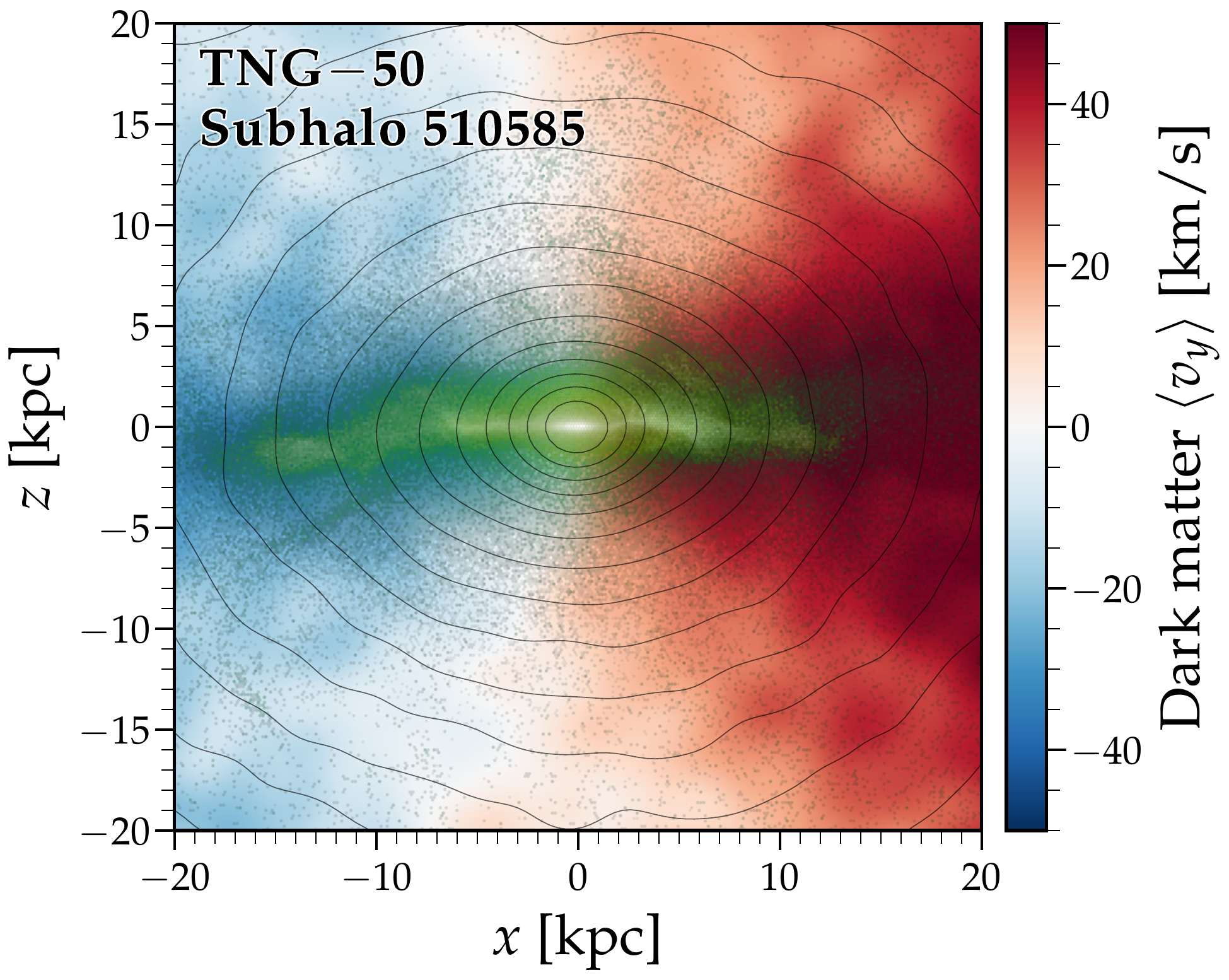}
    \includegraphics[height=0.42\linewidth,trim={1.8cm 0.0cm 0cm 0.0cm}, clip]{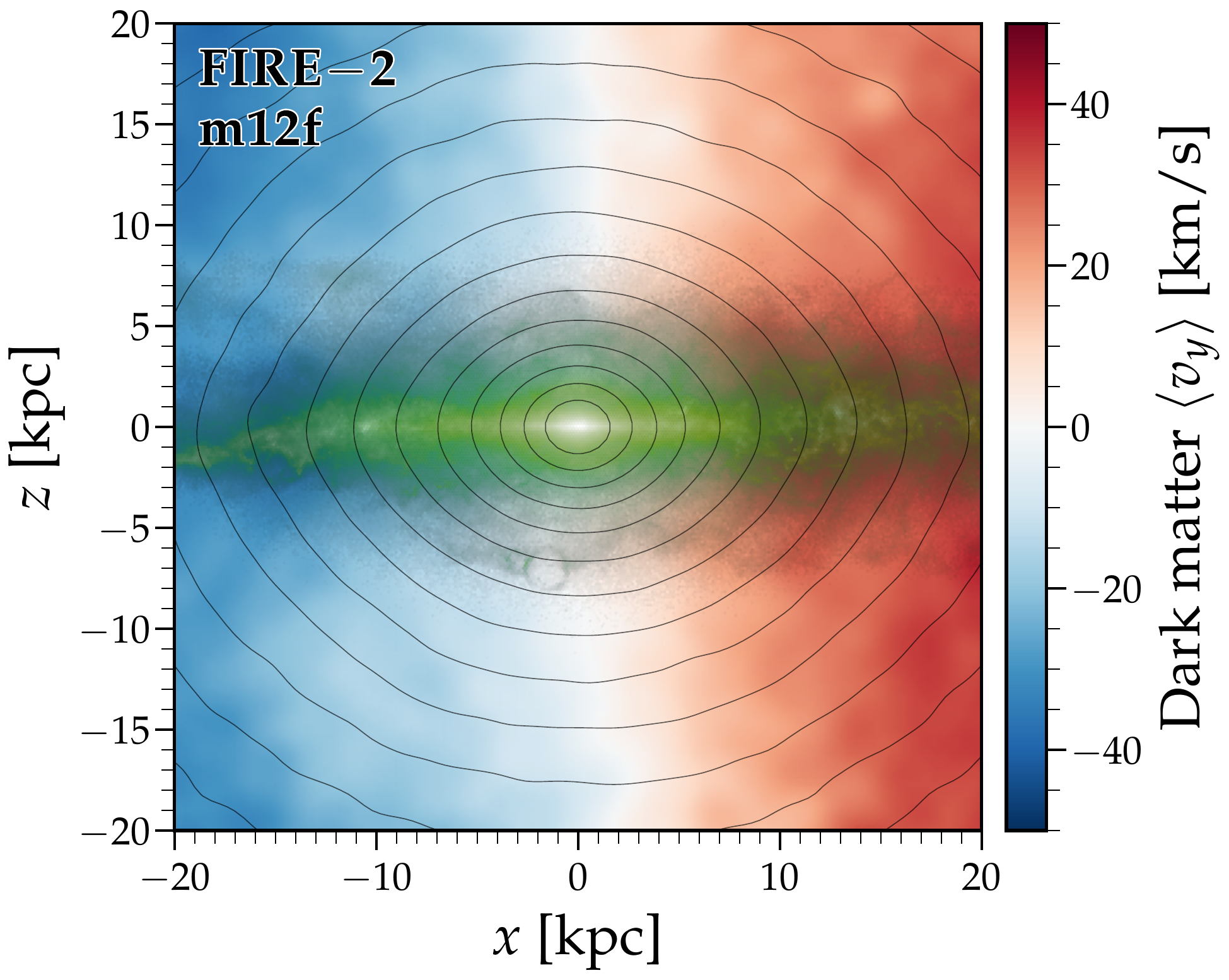}
    \caption{Two examples of Milky Way analogue galaxies used in this work. The left panel shows an analogue selected from IllustrisTNG-50 while the right panel shows a higher-resolution analogue from the FIRE-2/Latte zoom-in simulation. Gas and stars are shown in green, while the background colourmap indicates the per-pixel mean DM velocity along the $y$-coordinate (i.e.~into the page) to highlight the fact that these halos are co-rotating with the disk. The black lines are logarithmically spaced iso-density contours of the DM distribution projected along $y$.}
    \label{fig:snapshot}
\end{figure*}

Many direct searches for dark matter (DM) attempt to observe scattering events between particles incoming from our galactic halo and the nuclei or electrons inside a detection medium (see e.g.~\cite{MarrodanUndagoitia:2015veg, Schumann:2019eaa, Baudis:2025yva, Xia:2026bkh, Kahn:2021ttr, Lin:2022hnt, Cirelli:2024ssz} for reviews). As well as the details of the DM particle model in question, the signals predicted in these experiments depend critically on the galactic DM phase-space distribution in the Solar neighbourhood~\cite{Green:2011bv}. The simplest assumption for the Milky Way's DM halo is the Standard Halo Model (SHM), in which the local Galactic-frame velocity distribution, $f(\mathbf{v})$, is an isotropic Gaussian truncated at an escape speed. This remains the fiducial assumption used in interpretations of direct detection experiments~\cite{Lewin:1995rx,Baxter:2021pqo,Evans:2018bqy,ParticleDataGroup:2024cfk}, although numerous departures have been explored, including: triaxiality~\cite{Evans:2000gr}, velocity anisotropy~\cite{Bozorgnia:2013pua,Fornasa:2013iaa,Evans:2018bqy}, non-Maxwellian structure~\cite{Kuhlen:2009vh,McCabe:2010zh,Green:2010gw,OHare:2016pjy,Hryczuk:2020trm}, co-rotating DM components like dark disks~\cite{Billard:2012qu,Purcell:2009yp,Ling:2009cn,Bruch:2008rx,Folsom:2026dqs}, and kinematically cold substructures such as surviving subhalos~\cite{Baum:2021chx,Zhang:2025xzi} or debris from disrupted ones~\cite{Savage:2006qr,Lisanti:2011as,Kuhlen:2012fz,Purcell:2012sh,OHare:2014nxd,Kavanagh:2016xfi,OHare:2018trr,OHare:2019qxc,DEAP:2020iwi,Lawrence:2022niq,Maity:2022enp}. Since we do not have any methods to access the dark matter velocity distribution directly, numerous studies have sought to infer this quantity via N-body and hydrodynamic simulations of Milky Way-like galaxies, see e.g.~\cite{Moore:1999nt,Gnedin:2004cx,Hansen:2005yj,Diemand:2008in,Ling:2009eh,Vogelsberger:2008qb,Tissera:2009cm,Lisanti:2010qx,Mao:2013nda,Kuhlen:2013tra,Mao:2012hf,Butsky:2015pya,Zavala:2015neh,Bozorgnia:2016ogo,Kelso:2016qqj,Sloane:2016kyi,Bozorgnia:2017brl,OHare:2017yze,Necib:2018igl,Bozorgnia:2018pfa,Artale:2019wee,Bozorgnia:2019mjk,Poole-McKenzie:2020dbo,Nunez-Castineyra:2023dui,Staudt:2024tdq,Lilie:2025wkr}.

This work focuses on one particular issue: the structure of the high-speed tail of $f(\mathbf{v})$, i.e.~$|\mathbf{v}|\gtrsim 500$~km/s (in the galactocentric frame), which is subject to substantial observational uncertainty.\footnote{It is also possible for interactions between DM and other particles, e.g.~cosmic rays, to boost sub-populations of DM to relativistic or semi-relativistic speeds, e.g.~\cite{Bringmann:2018cvk, Ema:2018bih, Cappiello:2019qsw, Dent:2019krz, An:2017ojc, An:2021qdl, Emken:2017hnp, Granelli:2022ysi, Wang:2021jic, Agashe:2014yua, Cherry:2015oca, Bell:2026bhb, DeRocco:2019jti}. 
That is a model-dependent issue, which we are not discussing in this work.}

A sensible place to truncate $f(\mathbf{v})$ is at the local Galactic escape speed, $v_{\rm esc} = \sqrt{2\left(\Phi_{\infty}-\Phi_{\odot}\right)}$, where $\Phi_{\odot,\infty}$ are the values of the gravitational potential at the Solar position and at infinity, respectively. This quantity can in principle be inferred from nearby high-velocity stars, but such measurements are challenging and systematics-prone---results can be biased by the assumed prior on the functional form of the tail of the stellar velocity distribution; treatment of substructure, anisotropy, non-axisymmetry, and disequilibrium, all of which are present in our galaxy; as well as the mass modelling of the wider MW when going beyond local populations of stars~\cite{Grand:2019rma}. Even in the \textit{Gaia} era, measurements of $v_{\rm esc}$ at the solar position span the range~$v_{\rm esc}\approx 500$--$580$~km/s~\cite{Piffl:2014mfa,Monari_vesc,Williams_vesc,Deason:2019djb,Roche_vesc,Koppelman_vesc,Necib:2021vxr,Grand:2019rma,Wu_vesc}, all with large uncertainties (see Fig.~19 of Ref.~\cite{Hunt:2025uia} for a summary). Furthermore, translating a stellar escape-speed measurement into a statement about the DM distribution's high-velocity tail is non-trivial because the two populations have different global phase-space distributions, and because stars (and potentially DM particles) can be accelerated to high velocities by multi-body interactions. Additionally, recent simulations have suggested that the infall of the Large Magellanic Cloud (LMC) can introduce DM particles with speeds higher than the local escape speed into the Solar neighbourhood~\cite{Besla:2019xbx,SmithOrlik:2023lmc, ReynosoCordova:2024lmc, Garavito-Camargo:2020lqm}, which would open up a new kinematic regime for dark matter interactions~\cite{Bozorgnia:2025lsl, Reynoso-Cordova:2026pri}\footnote{Extragalactic DM particles extending to even higher speeds have also been discussed~\cite{Baushev:2012dm, Herrera:2023fpq, Santos-Santos:2023ubx, Kakharov:2025myy}.}, but without guidance from observations, precise statements remain difficult to make. In summary, the nature of the high-velocity tail of $f(\mathbf{v})$ is not well understood.

For many conventional elastic-scattering DM searches this uncertainty is often subdominant to others, because moderate changes in the high-speed tail---which is already exponentially suppressed---produce only modest changes in the total rate. However, the situation changes when a model is probed close to a kinematic threshold. This occurs if DM is a very light particle and so only the highest-speed particles deposit enough energy to be detected. It is also the case for endothermic inelastic DM, where the incoming particle must provide the energy required to access a heavier dark-sector state \cite{Tucker-Smith:2001myb,Tucker-Smith:2004mxa,Bramante:2016rdh,Barello:2014uda}. A well-motivated example that has received significant interest recently~\cite{Co:2021ion,Dessert:2022evk,Rodd:2024qsi,Abe:2025lci,Fan:2026kxx,Rodd:2026tyn,Freese:2026sga,Langhoff:2026ujr} is the Higgsino~\cite{Wells:2003tf,Nagata:2014wma}, which naturally explains the cosmological DM abundance as a thermal relic if its mass is $\sim 1.1$~TeV. This model has two neutral Majorana states that may be separated by a sub-MeV mass splitting and can therefore scatter inelastically via $Z$-boson exchange~\cite{Bottaro:2022one}.

A recent LZ analysis of an extended high-energy nuclear-recoil window\footnote{See also earlier analyses of this extended window~\cite{LZ:2023lvz,XENON:2022avm} which also discuss motivations for exploring it.} has brought this issue into focus again~\cite{LZ:2026axp}. With an exposure of $2.84~\mathrm{ton\,yr}$, LZ reported one event with characteristics consistent with a nuclear recoil at $E_R = 248 \pm 23_{\rm stat} \pm 23_{\rm sys}~\mathrm{keV}$.
The event lies in a region of negligible expected background. Numerous phenomenological explanations have already appeared which emphasise inelastic scattering as a possibility, e.g.~\cite{Yamashita:2026ump, Wu:2026nhi, Su:2026rwz, DiMauro:2026edge, Rodd:2026tyn, Freese:2026sga, Dent:2026bji, Nagata:2026pbj, Ge:2026xax, Xing:2026civ, Das:2026buc, Lian:2026hpm, Ahmed:2026kan, Lee:2026jxl,deLima:2026shq} (see also~\cite{Visinelli:2026kgt, Jeesun:2026vzo, Unwin:2026rdp, An:2026pkc, Arcadi:2026kev, Heikinheimo:2026kwp, Asadi:2026iot} for alternative explanations). Although a single event cannot establish a DM signal, its unusually high recoil energy motivates a re-assessment of astrophysical assumptions going into any DM interpretations.

\begin{figure*}[t]
    \centering
    \includegraphics[width=0.99\linewidth]{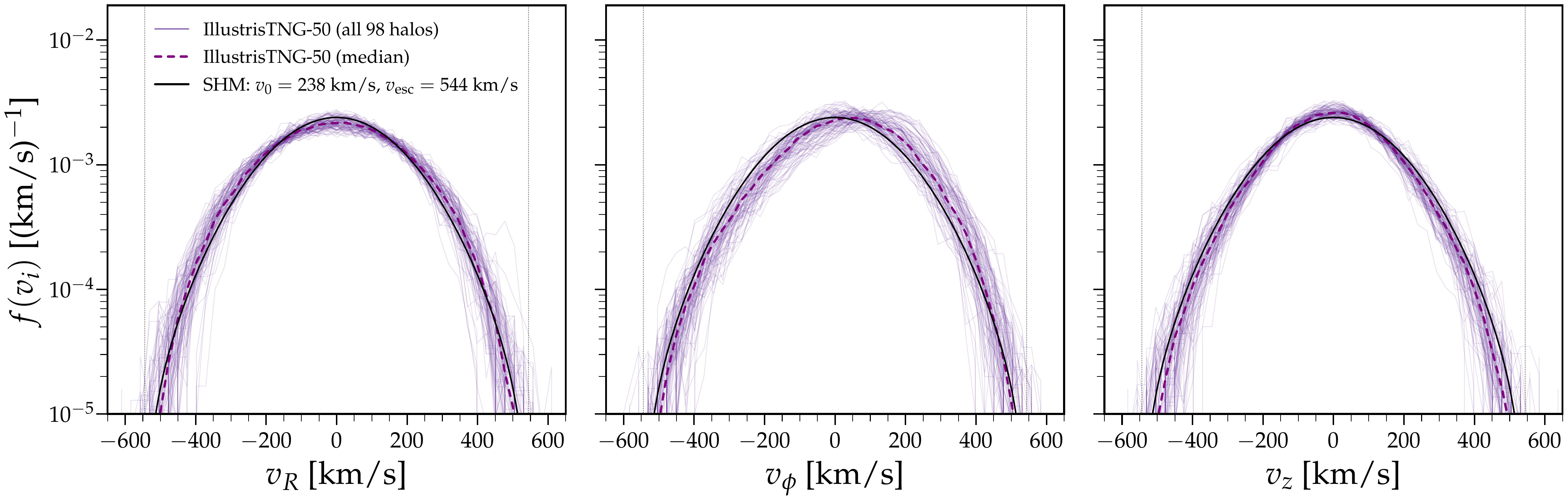}
    \caption{One-dimensional local velocity distributions in each cylindrical coordinate for DM particles in the IllustrisTNG-50 MW analogues. Individual analogues are shown with thin purple lines, while the median across all 98 halos is shown as a dashed line. The fiducial SHM is shown as a solid black line. The azimuthal component, $v_\phi$, exhibits the clearest departure from the SHM because of halo co-rotation.}
    \label{fig:fvi}
\end{figure*}

Here we use simulations to bound the range of possible high-speed DM populations. Inferring anything about a sparsely populated tail is naturally limited by that simulation's particle resolution, so here we combine a large set of 98 Milky Way (MW) analogues identified~\cite{Folsom:2024add} in the IllustrisTNG-50 hydrodynamical simulation~\cite{Pillepich:2019bmb} to provide additional coverage. For comparison, we also include three higher-resolution analogues from the Latte suite of the FIRE-2 cosmological zoom-in simulation~\cite{Wetzel2016, Hopkins:2017fire, Wetzel:2022man}.

An important feature of these simulated systems noted in Ref.~\cite{Folsom:2026dqs} is the mild \textit{co-rotation} of the DM halo with the baryonic disk, with a median azimuthal speed in the galactic plane around the solar neighbourhood in the range \mbox{$v_\phi \sim 6$--$70$~km/s}. Figure~\ref{fig:snapshot} shows a visual example of this rotating structure, which is seen in halos across both simulation suites. Co-rotation reduces the intensity of the DM wind seen by a terrestrial detector in the moving laboratory frame. This effect has the potential to reduce the maximum available energy that DM particles can supply to interactions inside a detector because the fastest particles are those at the tail of the galactic-frame velocity distribution and are travelling in the direction opposing the Earth's motion through the halo. Interestingly, however, as we show below, this co-rotation does not significantly affect the tail of the velocity distribution, because it skews the $f(v_\phi)$ distribution towards positive $v_\phi$ rather than simply shifting it to a non-zero mean.

The structure of this article is as follows: We first discuss the MW analogues we use in Sec.~\ref{sec:tng}, then outline generic implications of these simulations for scattering-based direct detection experiments in Sec.~\ref{sec:direct_detection}, before illustrating this in the context of an inelastic DM interpretation of the LZ event in Sec.~\ref{sec:lz}.

\section{Milky Way analogues in simulations}
\label{sec:tng}

A sample of 98 MW-like galaxies was identified by Ref.~\cite{Folsom:2024add} within the highest-resolution IllustrisTNG volume. TNG-50~\cite{Nelson:2019jkf,Pillepich:2019bmb,Nelson:2018uso} follows the coupled evolution of dark matter, gas, stars, and black holes in a cosmological volume of side length $51.7~\mathrm{Mpc}$. The MW analogues found in Ref.~\cite{Folsom:2024add} were selected to have stellar masses in the range $4\times10^{10}~M_\odot < M_\star < 7.3\times10^{10}~M_\odot$ and to satisfy isolation requirements intended to reproduce the broad Galactic environment. A subset have merger histories including events that are thought to have had a major role in shaping our galaxy, including the Gaia--Sausage--Enceladus (GSE) merger (see e.g.~\cite{Belokurov:2018vpl, Helmi:2018hnh, Grand:2020kuj, Helmi:2020otr, Hunt:2025uia}) and the ongoing interaction with the LMC (see e.g.~\cite{SmithOrlik:2023lmc, ReynosoCordova:2024lmc, Garavito-Camargo:2020lqm} and~\cite{Buch:2024ssx, Brooks:2025vcy} for related studies). 

As a cross-check, we also test a much smaller set of higher-resolution MW mass analogues from a cosmological zoom-in simulation: FIRE-2/Latte~\cite{Wetzel2016, Hopkins:2017fire, Wetzel:2022man}. The three publicly available halos we use from this project---m12m, m12i, m12f---have similar halo masses but varying assembly histories. See Ref.~\cite{Zhang:2026qnl} for a more in-depth discussion.

\begin{figure*}[t]
    \centering
    \includegraphics[width=0.49\linewidth]{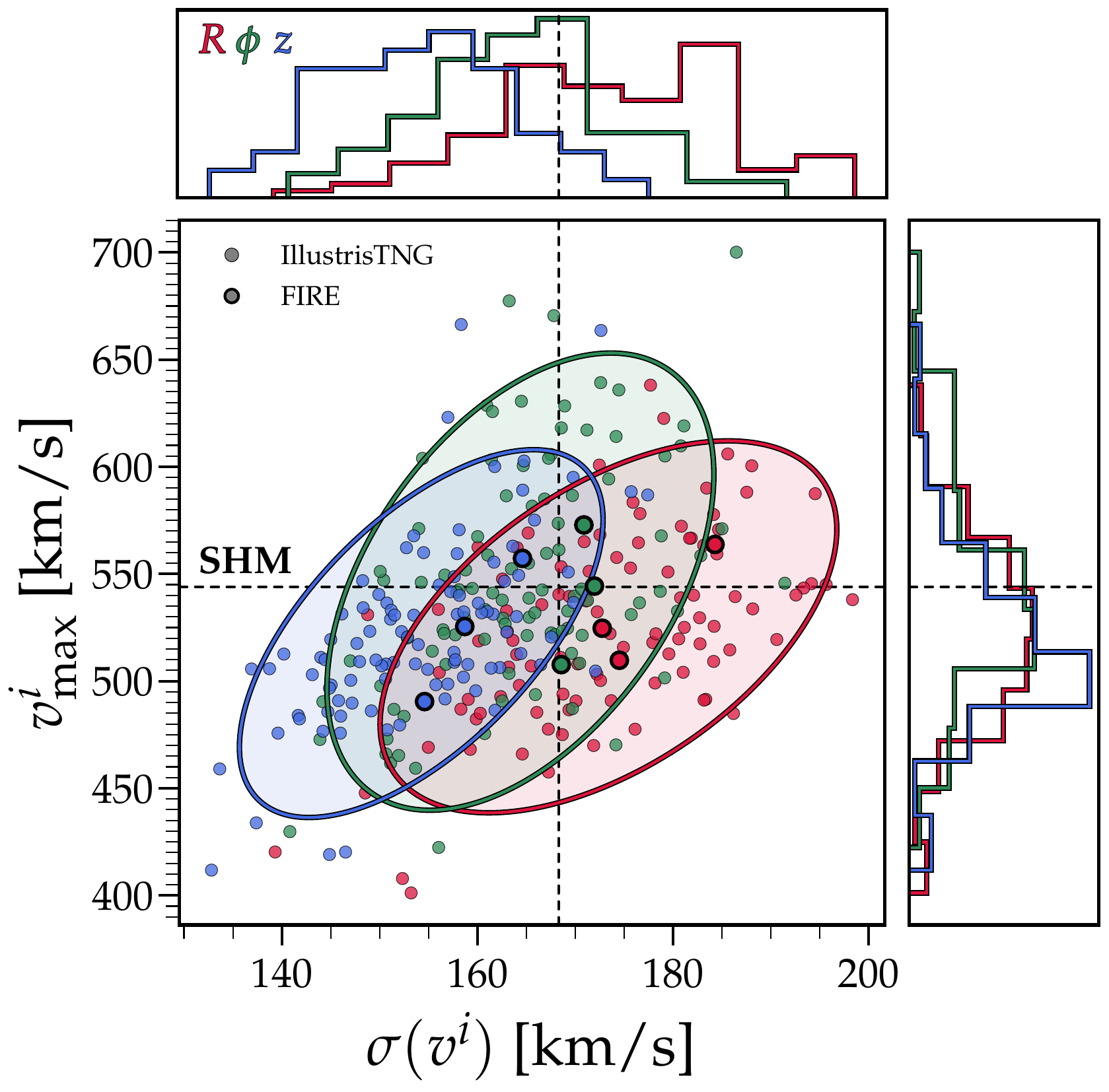}
    \includegraphics[width=0.49\linewidth]{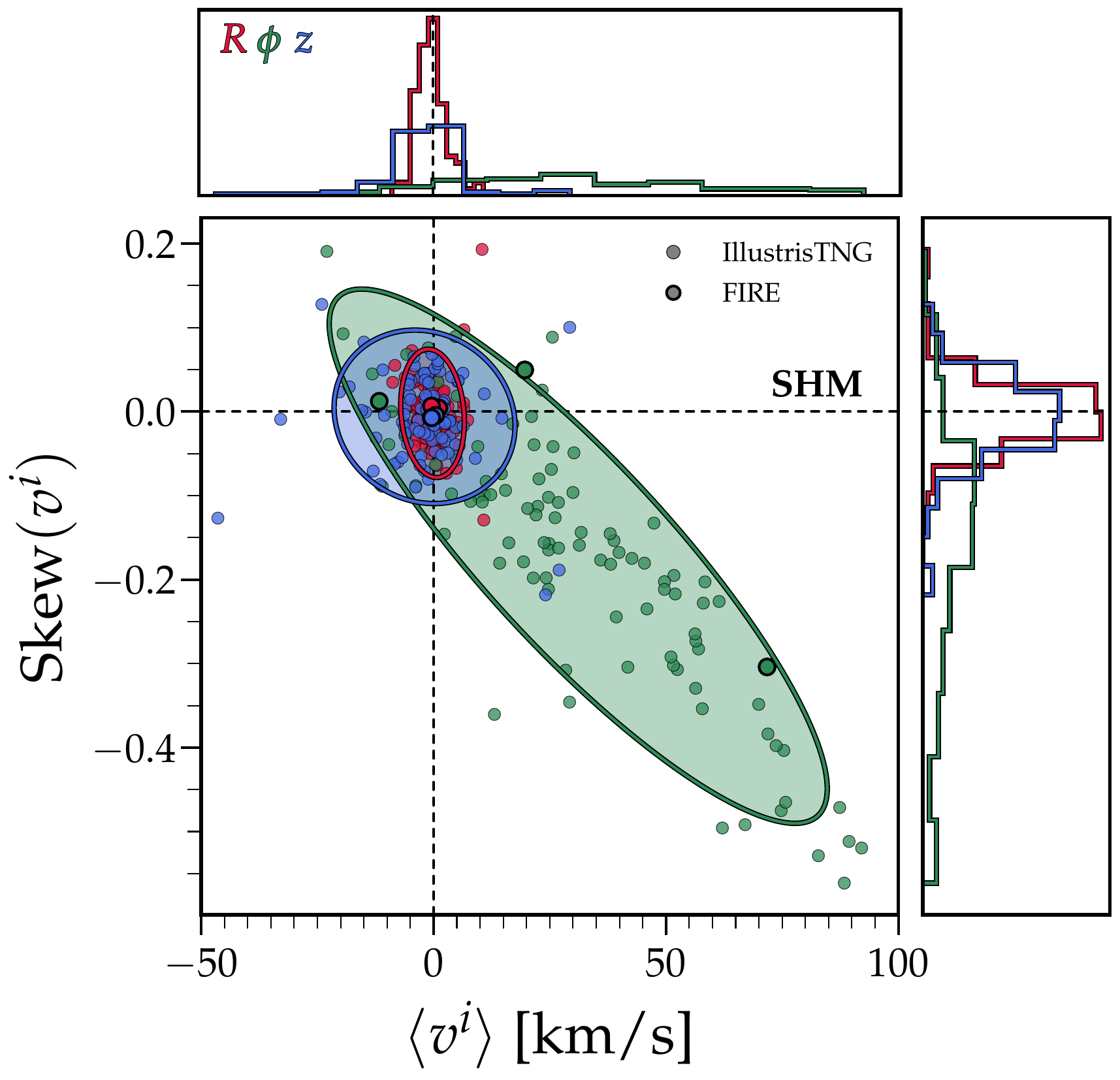}
    \caption{Summary statistics for the one-dimensional velocity distributions in each cylindrical coordinate ($R,\phi,z$ are red, green, blue, respectively). The left-hand panel shows the maximum speed ($v_{\rm max}^i \equiv \max|v_i|$) versus the standard deviation of the velocity in each coordinate and for each halo, while the right panel shows the skewness and the mean. The generic co-rotation of the halos manifests in this plot as a preference for $v_\phi$ to have a positive mean and negative skewness. We fit Gaussian $2\sigma$ contours to the distribution to provide a visual aid. The majority of the points are for the set of 98 Milky Way analogues from IllustrisTNG-50 studied in this work, but for comparison we also show three high-resolution analogues from FIRE (m12f, m12m, m12i), for which we see broadly consistent conclusions. Notably, m12f (visualised in Fig.~\ref{fig:snapshot}) shows a strong negative skew and has $\langle v_\phi \rangle \sim 70$~km/s.}
    \label{fig:summary_stats}
\end{figure*}

A direct comparison between simulated systems and the MW is complicated by differences in disk scale radius and rotation curve. The TNG-50 analogues are not identical to the MW and typically possess disks that are less centrally concentrated. It is therefore not straightforward to place an analogue Solar System in an analogue MW galaxy without making an arbitrary choice. For example, an orbital radius of $R_\odot = 8.3$~kpc in our galaxy is not necessarily equivalent to the same radius in another galaxy with a different stellar density profile, because the enclosed mass and hence the circular rotation speed will also differ. Ref.~\cite{Folsom:2025lly} proposed a more principled approach that uses a simple energy-conserving rescaling to compress each simulated galaxy in position and velocity space. The aim of this technique is to generate particle distributions at the Solar circle for which the local standard of rest (the velocity of a circular orbit passing through our location) is matched to its well-measured value in our galaxy.

We follow this prescription and adopt \mbox{$v_{\rm LSR}=238~\mathrm{km\,s^{-1}}$} (see e.g.~\cite{Bobylev_LSR, Zbinden_LSR, Ding_LSR, Gaia_accel, Hunt:2025uia}) for the local standard of rest, which is determined to $\sim$5~km/s precision, thanks to precise measurements of the proper motion and distance of Sgr A*~\cite{Bland-Hawthorn:2016lwg, Reid:2004rd, GRAVITY2021} and the Sun's peculiar velocity away from a perfectly circular orbit~\cite{Schoenrich_vpec}.

Our procedure is as follows. We extract the star, gas, and DM particle data for the 98 TNG-50 halos~\cite{Nelson:2019jkf, Pillepich:2019bmb, Nelson:2018uso} and three FIRE-2/Latte halos; rotate each system so that the $z$ axis is aligned with the angular-momentum vector of the stellar distribution within twice the stellar half-mass radius; apply the phase-space rescaling described in Ref.~\cite{Folsom:2025lly} to match the local standard of rest at the Solar circle; and finally define a ``local'' population of DM particles to be those located inside a torus of radius $R_\odot=8.3~\mathrm{kpc}$~\cite{GRAVITY2021} with cross-sectional radius $r=1~\mathrm{kpc}$. Our conclusions are unchanged when we vary $r$ by $0.5~\mathrm{kpc}$ up or down. We marginalise over the arbitrary azimuth of the Solar position by expressing the velocity distribution of local DM particles in cylindrical coordinates ($R,\phi,z$) where $z$ aligns with the angular momentum of the galactic disk.

Figure~\ref{fig:snapshot} shows side-on views of two representative systems from TNG-50 and FIRE, before rescaling. The fact that these systems are both ``MW analogues'' but have baryonic disks with very different structures and sizes is part of the problem that the scaling procedure aims to resolve.

In Fig.~\ref{fig:snapshot} we also show the DM velocity field. It can be seen by eye that these DM halos are spinning with $v_\phi \sim \mathcal{O}(10\,{\rm km/s})$, and that this spin roughly aligns with the angular momentum of the baryonic disk. This systematic co-rotation was noted in Ref.~\cite{Folsom:2026dqs} for the TNG-50 halos and in Ref.~\cite{Zhang:2026qnl} for FIRE. This observation has precedent in much earlier simulation literature~\cite{Read:2008fh, Read:2009iv, Purcell:2009yp}, and it may relate to the phenomenon referred to as a dark disk---we note that the DM density profiles shown in Fig.~\ref{fig:snapshot} are triaxial and flattened along the $x-y$ plane. The specific reasons behind the preference for co-rotation in these simulations have not been fully investigated, but figure rotation of triaxial DM halos is a generic expectation in $\Lambda$CDM as a result of torques exerted during their hierarchical assembly; see e.g.~\cite{Ash:2026jpj} for a recent discussion. If a significant stellar and gas disk forms early, then it is possible for incoming satellites to feel this potential and merge preferentially along the disk plane, leading the halo to accumulate angular momentum in a common direction.

Using the population of ``local'' DM particles defined above, we can construct their velocity distributions. These are generally expected to lie close to the SHM's truncated Gaussian velocity distribution, but with some notable differences. The fiducial SHM velocity distribution is written as,
\begin{equation}
    f_{\rm SHM}(\mathbf{v}) \propto
    e^{-|\mathbf{v}|^2/v_0^2}\,\Theta(v_{\rm esc}-|\mathbf{v}|),
\end{equation}
with $v_0=238~\mathrm{km\,s^{-1}}$ and $v_{\rm esc}=544~\mathrm{km\,s^{-1}}$ \cite{Baxter:2021pqo}. We obtain the corresponding one-dimensional velocity distributions along each cylindrical coordinate, $f(v_R)$, etc., by integrating over the other two velocity components.

Figure~\ref{fig:fvi} compares the one-dimensional velocity distributions in the TNG-50 halos alongside the fiducial SHM. The radial and vertical components, $f(v_R)$ and $f(v_z)$, are centred on zero, as expected, but are slightly hotter and colder on average than the SHM, respectively. All three distributions have excess kurtosis, which is particularly noticeable in the $f(v_z)$ case. The azimuthal component, $f(v_\phi)$, meanwhile, shows a clear preference for positive velocities, as described above, with a mean of $\langle v_\phi \rangle \sim 5-58$~km/s (68\% range). However, this co-rotation does not noticeably affect the tails of the distribution, which remain in roughly the same range across all halos and between all three coordinates. Rather than $f(v_\phi)$ being shifted towards positive $v_\phi$, the distribution is skewed, with a longer tail for negative $v_\phi$.

We illustrate this behaviour more clearly in Fig.~\ref{fig:summary_stats}. We collapse the three one-dimensional velocity distributions into summary statistics---the maximum unsigned velocity component, standard deviation, skewness, and the mean---and plot these for each halo. We again see that the halos are generally hottest in the radial coordinate ($\sigma(v_R)$ is largest). This is reminiscent of what is expected in our galaxy, where the GSE merger event is believed to have been a head-on collision that brought debris into our halo on eccentric, radial orbits. With the exception of the $v_\phi$ distributions, which have pronounced negative skew, the summary statistics are otherwise generally consistent across halos and across the three coordinates. The median of the halo-to-halo distribution is also very close to the SHM, but this is partly by construction because the scaling procedure makes it so that all the halos have the same local standard of rest as the fiducial SHM.

The three FIRE-2 systems we include occupy broadly similar regions of this summary-statistic space despite their differences in simulation techniques, handling of baryonic physics, and the assembly histories of the resulting analogues. In particular, the m12f halo exhibits a skewed $v_\phi$ that is close to some of the more extreme co-rotating TNG-50 halos.

\section{Implications for direct dark matter detection}
\label{sec:direct_detection}

\begin{figure}[!t]
    \centering
    \includegraphics[width=\linewidth]{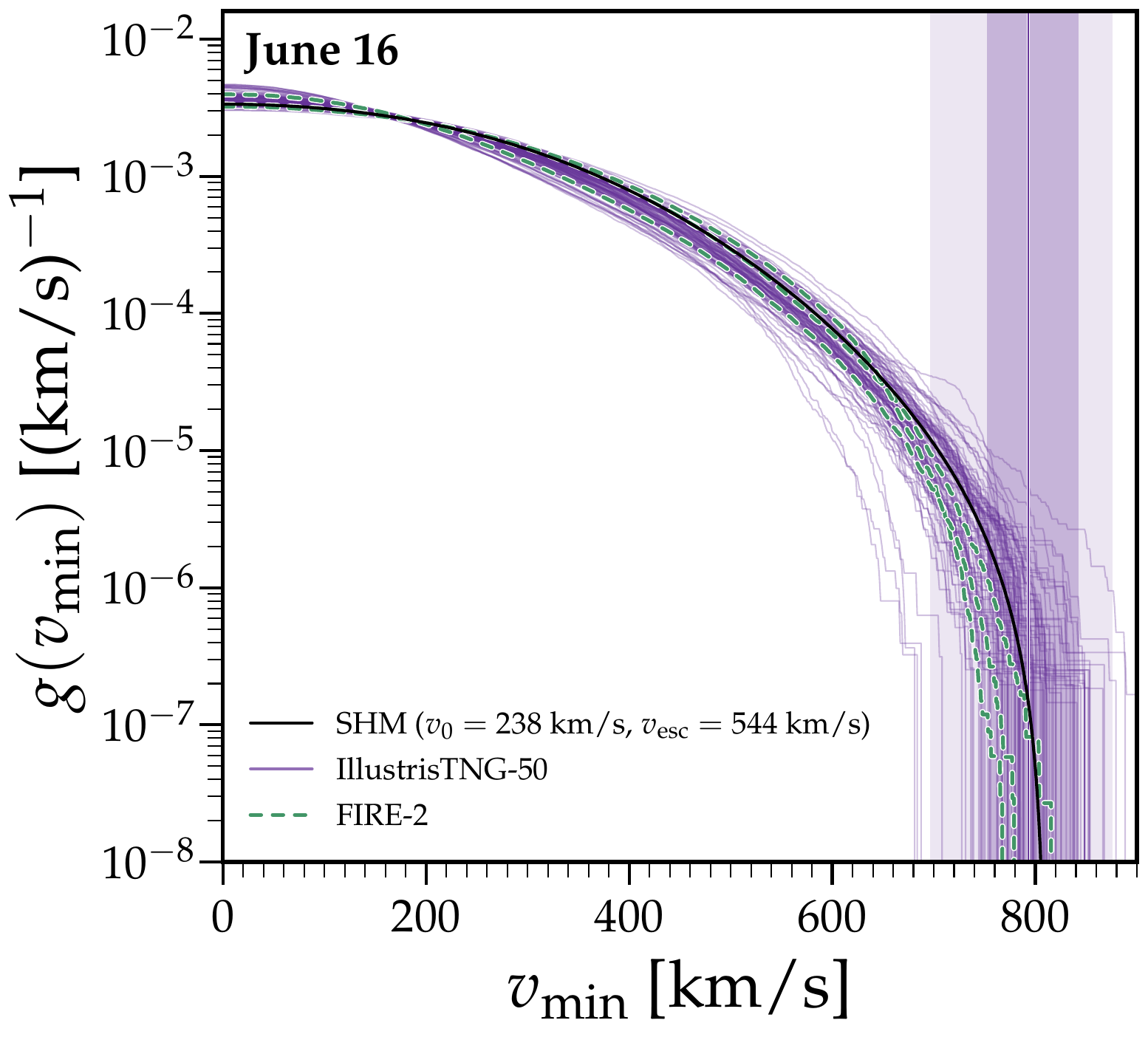}
    \caption{The mean inverse speed of dark matter particles with laboratory-frame speeds $v>v_{\rm min}$ as a function of $v_{\rm min}$. The laboratory velocity is taken at its value on June 16. All 98 of the TNG-50 halos are shown as purple lines, and the three FIRE-2 halos are shown as green dashed lines. The 68\% and 95\% containment of the maximum laboratory-frame dark matter speed on June 16 are shown with a vertical band. The median across all halos is almost identical to the fiducial SHM shown as a black line.}
    \label{fig:gvmin}
\end{figure}

\begin{figure}[t]
    \centering
    \includegraphics[width=\linewidth]{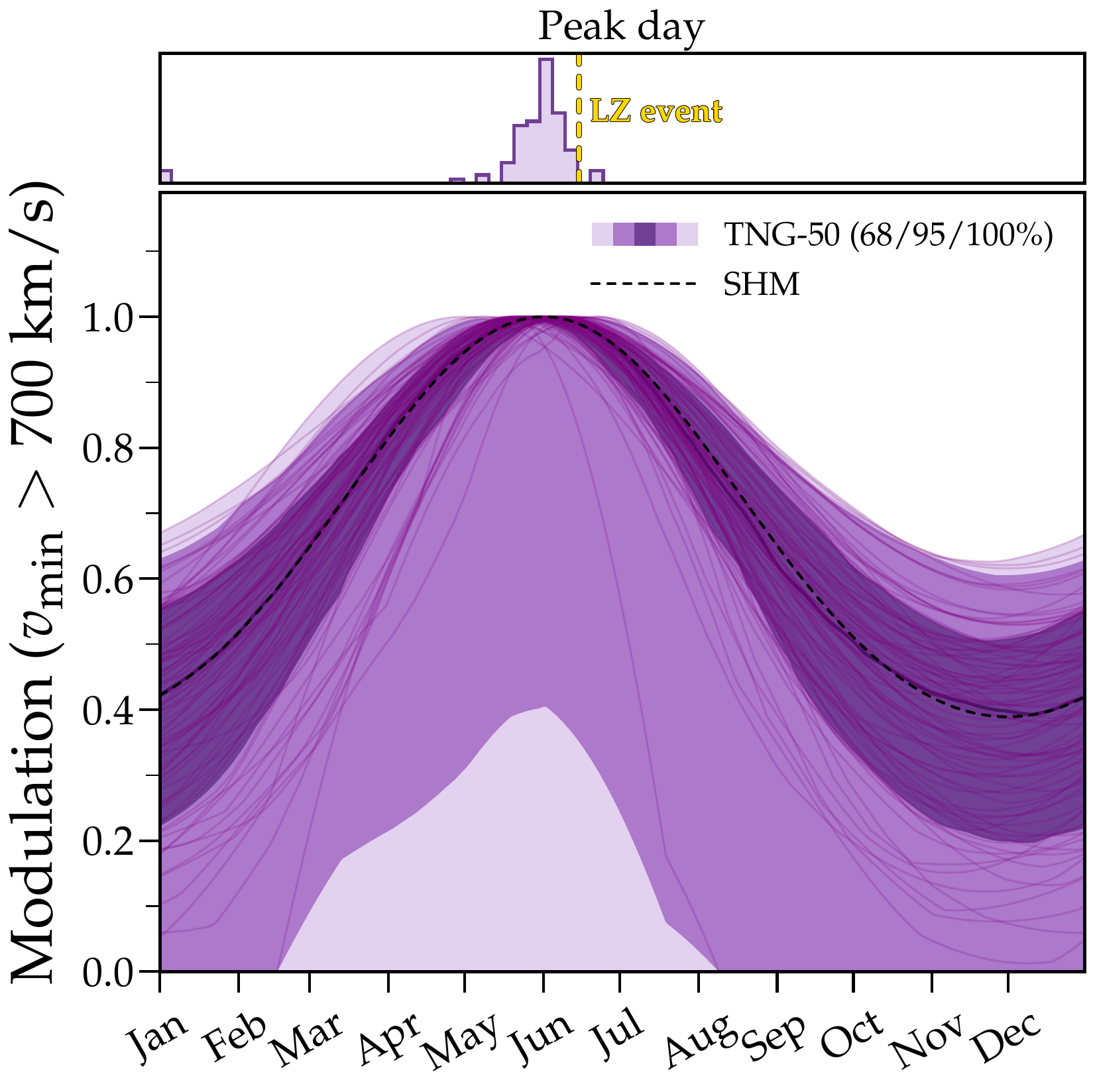}
    \caption{Annual modulation of a high-threshold signal, illustrated by integrating the mean inverse speed for $v_{\min}>700~\mathrm{km\,s^{-1}}$ as a function of time. We normalise this integral by the maximum value it achieves over the year. Each thin purple line corresponds to one simulated halo, while coloured bands contain 68, 95, and 100\% of simulations. The black dashed line is the fiducial SHM. Above the main plot, we show a histogram of the day when the modulation is maximised.}
    \label{fig:modulation}
\end{figure}

We now take the full range of simulated velocity distributions and propagate this uncertainty into the signals in dark matter experiments. We will focus on the generic DM scenario exploited by the largest detectors, in which DM is assumed to scatter on nuclei in some way. For DM-nucleus scattering, the differential rate per unit detector mass can be written as,
\begin{equation}
    \frac{{\rm d}R(t)}{{\rm d}E_R}
    =
    \frac{\rho_\chi}{m_\chi}
    \sum_T
    \frac{\xi_T}{m_T}
    \int_{v>v_{\min}(E_R)}
    {\rm d}^3v\,
    f_{\rm lab}(\mathbf{v},t)\,
    v\,
    \frac{{\rm d}\sigma_T}{{\rm d}E_R}\,,
    \label{eq:generalrate}
\end{equation}
where $\rho_\chi\approx 0.4~\mathrm{GeV\,cm^{-3}}$ is the local DM density~\cite{Read:2014qva,deSalas:2020hbh,Putney:2025mch}, $T$ labels target isotopes, $\xi_T$ are their corresponding mass fractions, and $f_{\rm lab}$ is the laboratory-frame velocity distribution.

For interactions with a velocity-independent matrix element, we have ${\rm d}\sigma/{\rm d}E_R\propto v^{-2}$, so all dependence on the velocity distribution is contained in one function, the mean inverse speed,
\begin{equation}
    g(v_{\min},t)
    =
    \int_{v>v_{\min}}
    \frac{f_{\rm lab}(\mathbf{v},t)}{v}\,{\rm d}^3v\,.
    \label{eq:eta}
\end{equation}
For example, the canonical spin-independent interaction with equal neutron and proton couplings yields the following simple expression for the rate,
\begin{equation}
    \frac{dR(t)}{dE_R}
    =
    \frac{\rho_\chi\,\sigma_n}{2m_\chi\mu_{\chi n}^2}
    \sum_T
    \xi_T A_T^2
    F_T^2(E_R)\,
    g\!\left[v_{\min}(E_R),t\right]\,,
    \label{eq:sirate}
\end{equation}
where $\sigma_n$ is a reference DM--nucleon cross section, \mbox{$\mu_{\chi n} = m_\chi m_n/(m_\chi +m_n)$} is the DM--nucleon reduced mass, and $F_T(E_R)$ are form factors that describe the nuclear structure of isotope $T$.

The important quantity for this discussion is the laboratory-frame velocity distribution. This is obtained by boosting the Galactic-frame distribution, $f(\mathbf{v})$, by the velocity of the laboratory with respect to the Galactic centre:
\begin{equation}
    f_{\rm lab}(\mathbf{v},t)
    =
    f\!\left(\mathbf{v}+\mathbf{v}_{\rm lab}(t)\right).
\end{equation}
The boost velocity, $\mathbf{v}_{\rm lab}(t) = \mathbf{v}_\odot + \mathbf{v}_\oplus(t)$, is the sum of the orbital motion of the Sun around the galaxy, \mbox{$\mathbf{v}_\odot = (-11.1, 250.24,   7.25)$}~\cite{Schoenrich_vpec} and the orbital motion of the Earth with respect to the Sun (see e.g.~\cite{McCabe:2013kea, Mayet:2016zxu}). Ignoring corrections from the Earth's orbital eccentricity, the latter can be written as,
\begin{equation}
\mathbf{v}_{\oplus}(t)
=
v_{\oplus}
\left[
\cos\!\left[\omega(t-t_a)\right]\hat{\boldsymbol{\epsilon}}_1
+
\sin\!\left[\omega(t-t_a)\right]\hat{\boldsymbol{\epsilon}}_2
\right],
\label{eq:earth_velocity}
\end{equation}
where $v_\oplus=29.79~\mathrm{km\,s^{-1}}$, $\omega=2\pi/(1\,\mathrm{yr})$, $t_a\simeq22$ March, and,
\begin{align}
\hat{\boldsymbol{\epsilon}}_1 &= (-0.9941,\,0.1088,\,0.0042),\\
\hat{\boldsymbol{\epsilon}}_2 &= (0.0504,\,0.4946,\,-0.8677),
\end{align}
form an orthonormal basis for the Earth's orbital plane in Galactic coordinates\footnote{Note that we have applied a minus sign to the first coordinate in $\mathbf{v}_\oplus$ and $\mathbf{v}_\odot$ since we are working in a $(v_R,v_\phi,v_z)$ system instead of the Cartesian $(U, V, W)$ system.}. Because this component is time-dependent, $g(v_{\rm min},t)$ and $\textrm{d}R(t)/\textrm{d}E_R$ also pick up a characteristic yearly oscillation known as annual modulation~\cite{Freese:2012xd, Froborg:2020tdh}.

We now show how uncertainties in $f(\mathbf{v})$ impact DM-scattering experiments by computing the mean inverse speed $g(v_{\min},t)$ for our simulated halos. Since $g(v_{\rm min})$ is effectively a cumulative distribution, we can compute it directly using simulation particles without having to bin them or find an analytic fit. Figure~\ref{fig:gvmin} shows the resulting $g(v_{\rm min})$ evaluated on June 16---this is the date the LZ event was observed~\cite{LZ:2026axp}, and is close to the day of the year when $|\mathbf{v}_{\rm lab}|$ is maximised. 

At moderate $v_{\min}$, the resulting $g(v_{\rm min},t={\rm June~16})$ curves for our simulations cluster closely around the SHM, but are slightly larger than the SHM for $v_{\rm min}<200$~km/s and smaller above this value. This weighting towards lower speeds is again due to the co-rotation of these halos, which reduces lab-frame DM speeds on average. Above $700~\mathrm{km\,s^{-1}}$, close to the particle resolution of the simulations, the halo-to-halo variation in the tail becomes dominant. Across the TNG-50 sample, the maximum laboratory-frame speed on June 16 is $v_{\rm max}^{\rm lab} = 792.7^{+82.9}_{-95.6}~\mathrm{km\,s^{-1}}$ (median and 95\% containment), also shown as a vertical band in Fig.~\ref{fig:gvmin}.

The annual modulation effect can be greatly amplified in DM models where only events above a large value of $v_{\rm min}$ are observed. For example, if the threshold $v_{\rm min}$ is larger than 700 km/s, the amplitude of annual modulation could be as large as 100\%, i.e.~the event rate drops to zero for part of the year~\cite{McCabe:2026crm}. We illustrate this in a DM-model-independent way in Fig.~\ref{fig:modulation} by integrating just the high-speed contribution: $g_{\rm high}(t) = \int_{\rm 700\,{\rm km/s}}^\infty \textrm{d}v_{\rm min}g(v_{\rm min},t)$ and plot this as a relative modulation over the year, i.e.~$M(t) = g_{\rm high}(t)/{\rm max}(g_{\rm high})$.

We see that most of our halos still show a maximum in the event rate around June, but with a spread of about a month. On average, the halos have an almost identical modulation to the fiducial SHM, but with a subset of cases having a much sharper seasonal switching-on of the event rate, due to the slight preference for suppressed high-speed tails. In extreme cases ($\sim$5\% of halos), the event rate can almost vanish for three-quarters of the year even with this modest threshold value of $v_{\rm min}$.

\section{The LZ event and inelastic dark matter}
\label{sec:lz}

\begin{figure}[t]
    \centering
    \includegraphics[width=\linewidth]{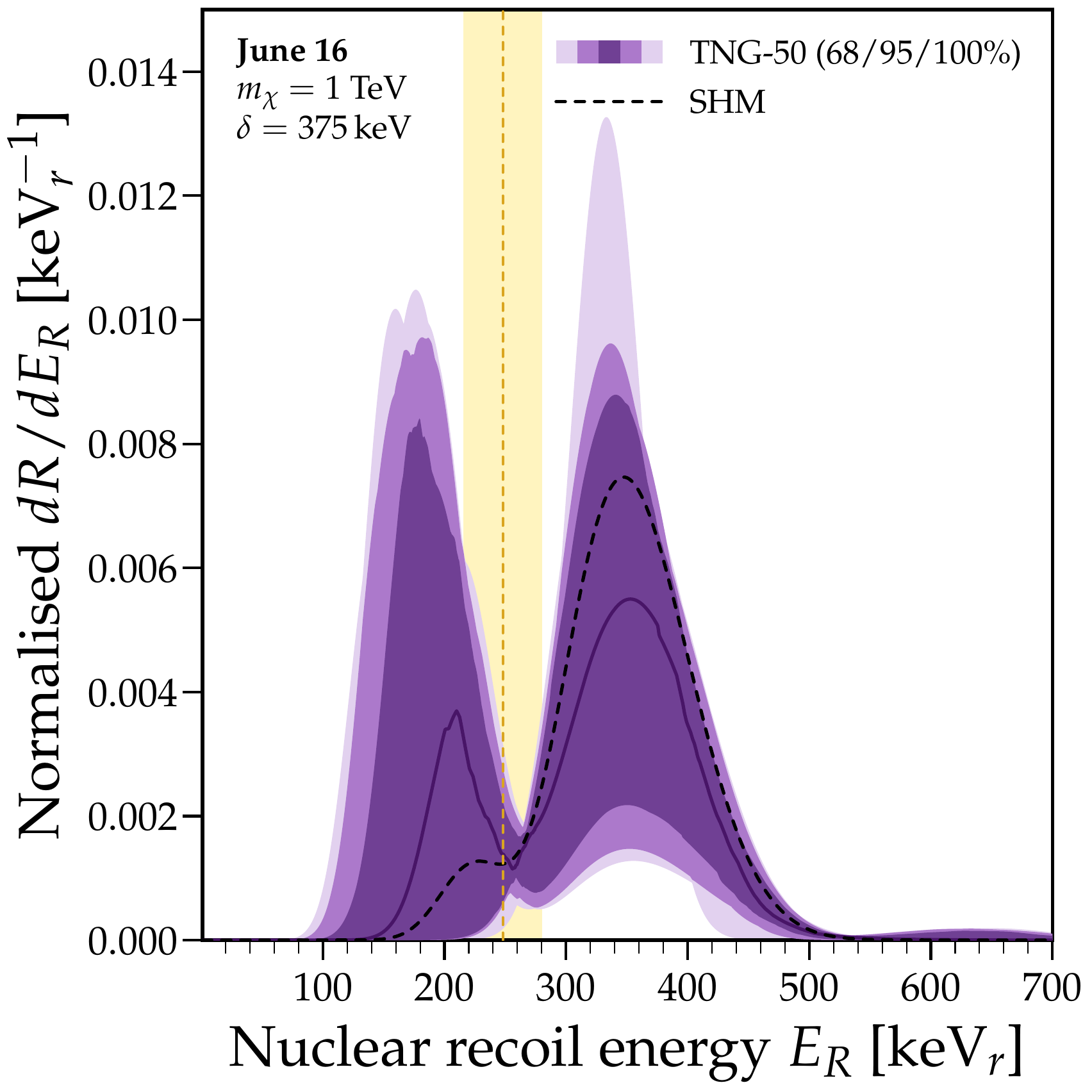}
    \caption{Normalised recoil spectra for endothermic inelastic DM with $m_\chi=1~\mathrm{TeV}$ and $\delta=375~\mathrm{keV}$ on xenon. The shaded bands show the halo-to-halo spread obtained from the simulated $g(v_{\min})$ functions, while the SHM prediction is shown for comparison. The measured LZ recoil energy and its uncertainty are indicated by the gold vertical band. The spectra are normalised to unit area to isolate the effect of the velocity distribution on the spectral shape as opposed to the total event rate.}
    \label{fig:drde}
\end{figure}

\begin{figure}[t]
    \centering
    \includegraphics[width=\linewidth]{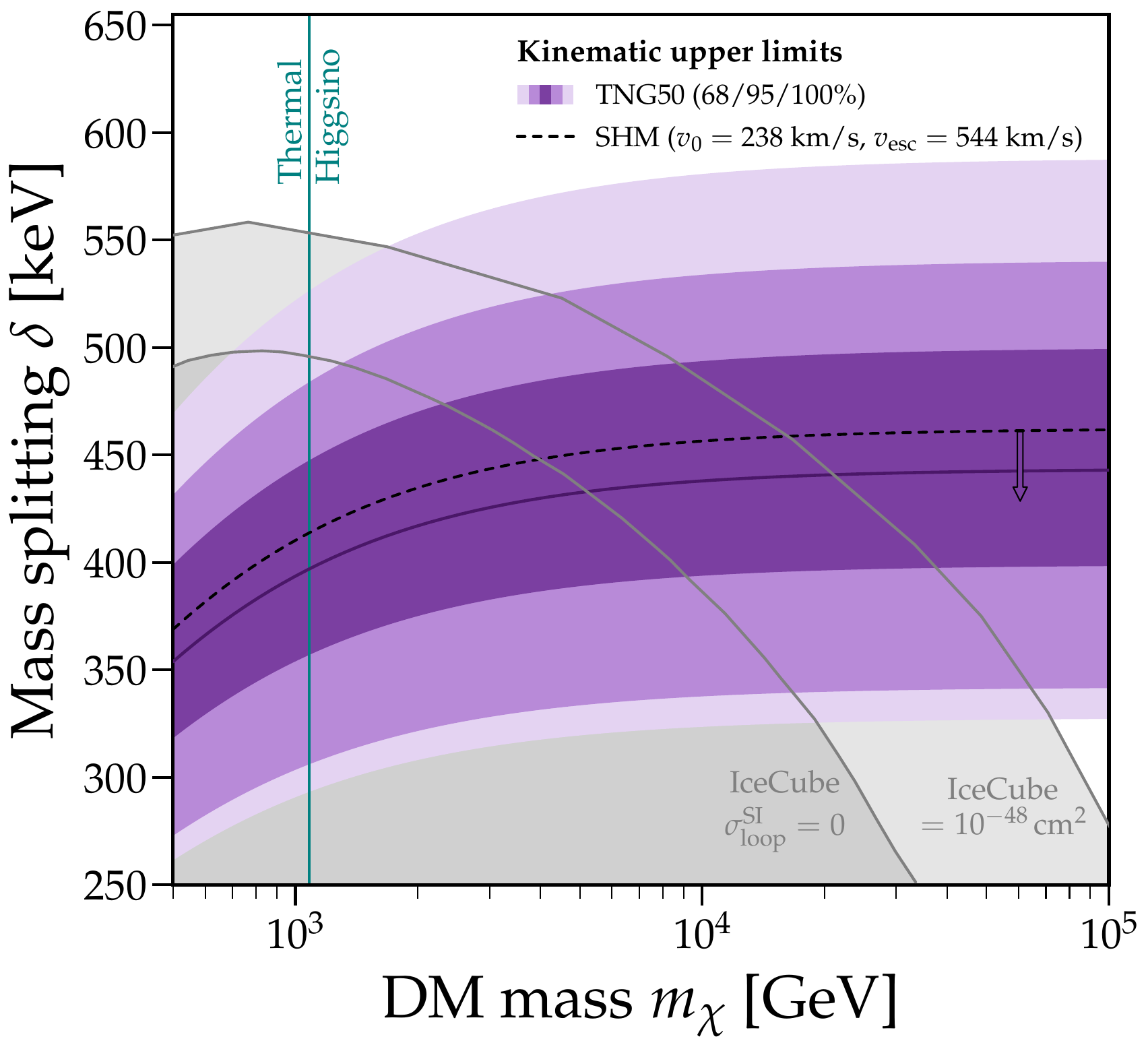}
    \caption{Kinematic upper limit on the inelastic splitting $\delta$ as a function of DM mass, obtained by requiring scattering on xenon to remain accessible for the largest laboratory-frame speeds. The spread reflects the halo-to-halo variation of the maximum DM speed. The fiducial SHM limit is shown as a black dashed line, while the median across all halos is shown as a solid purple line. We also show solar-capture constraints on the Higgsino model from IceCube taken from Refs.~\cite{Bose:2026ndd} for two different choices of the loop-level DM-nucleon coupling (see also \cite{Pospelov:2026ewn,Nguyen:2026lui}). Any endothermic inelastic DM explanation of the LZ event must lie below one of the purple lines, while IceCube bounds the Higgsino case to be above the grey region.}
    \label{fig:kinematic}
\end{figure}

The LZ Collaboration recently reported a search for nuclear recoil events extending up to $270~\mathrm{keV}$, collected over a $2.84~\mathrm{ton\, yr}$ exposure~\cite{LZ:2026axp}. One event consistent with the properties of a nuclear recoil was observed in a region of no known backgrounds. If the event is interpreted as a nuclear recoil, its energy would be: $E_R = 248 \pm 23_{\rm stat} \pm 23_{\rm sys}~\mathrm{keV}$, with uncertainties stemming from the conversion of the measured charge and light signals into the kinetic energy imparted to the xenon nucleus by the interaction.

LZ observed only this one high-energy recoil and not the canonical exponentially falling tail of recoils expected from most DM models, which generate nuclear recoils via elastic scattering. The event can be naturally explained if DM scatters inelastically, because in this case, lower-energy recoils would be kinematically forbidden. Inelastic DM models involve the scattering process,
\begin{equation}
    \chi_1 + N \rightarrow \chi_2 + N,
\end{equation}
with,
\begin{equation}
    \delta \equiv m_{\chi_2}-m_{\chi_1}.
\end{equation}
With this convention, $\delta>0$ denotes endothermic upscattering and $\delta<0$ exothermic down-scattering~\cite{Graham:2010ca}. We will take the former example to showcase the important role of the high-velocity tail, though both scenarios can explain the LZ event~\cite{Dent:2026bji}.

For a nucleus of mass $m_T$, the minimum incident speed required to produce a recoil with energy $E_R$ is then,
\begin{equation}
    v_{\min}(E_R)
    =
    \frac{1}{\sqrt{2m_T E_R}}
    \left|
        \frac{m_T E_R}{\mu_{\chi T}}+\delta
    \right|,
    \label{eq:vmin}
\end{equation}
where $\mu_{\chi T}=m_\chi m_T/(m_\chi+m_T)$. Positive $\delta$ raises the speed threshold required to produce recoil events, and depending on the choice of $m_\chi$, a recoil near $248~\mathrm{keV}$ may require $v_{\min}$ close to the highest lab-frame DM speeds provided by the Galactic halo. The predicted event rate can also change by orders of magnitude under comparatively modest changes in the tail.

To illustrate this point, we compute the recoil energy spectrum for an endothermic inelastic DM model scattering on xenon, using our simulated $g(v_{\rm min})$ curves derived in the previous section.

We will assume, for concreteness, that the DM scatters with equal couplings on neutrons and protons. Since we focus on the shape of the spectrum, we leave the absolute cross section arbitrary. We take spin-independent structure factors calculated using shell-model methods by Ref.~\cite{Vietze:2014vja} and sum over the stable xenon isotopes.

Figure~\ref{fig:drde} shows an example recoil spectrum for a $(m_\chi,\delta)$ parameter set able to explain the LZ event: $m_\chi=1~\mathrm{TeV}$ and $\delta=375~\mathrm{keV}$. We convolve the recoil spectrum with a Gaussian energy resolution kernel, which scales as $\sigma_E \propto \sqrt{E_R}$ and is normalised to match LZ's quoted uncertainty on the nuclear recoil event. We see that the recoil spectrum can be substantially distorted in this model because of the range of variation in the tails of the velocity distribution.

This particular choice of parameters has been intentionally chosen to align with thermal relic Higgsino dark matter (at $1.1$~TeV Higgsinos are thermally produced at the observed cosmological DM abundance~\cite{Bottaro:2022one}). However, this model is in tension with indirect bounds from IceCube (discussed further below) and potentially LZ itself, due to the large event rate predicted in LZ's empty higher-energy sideband at $\sim$350–680 keV~\cite{Rodd:2026tyn,Dent:2026bji}. Comparing the solid and dashed lines, we see that when averaging over all our halos, the predicted event rate spectrum is enhanced at energies smaller than the observed event, while the spectrum extending into LZ's high-energy sideband is suppressed. That said, with only one event to work with, this matter is difficult to settle conclusively until more data is collected by LZ and other experiments~\cite{XENON:2024wpa,PandaX:2018wtu,Baer:2026yrt}. The high-energy sideband is also uncalibrated for nuclear recoils, making it challenging to interpret any events seen in this region (or lack thereof).

Finally, we discuss the extent to which uncertainty in the highest DM speeds can be used to escape bounds on inelastic DM models. In particular, it was noted recently that the capture of this same DM candidate by the Sun, and subsequent annihilation into neutrinos (via $W^+W^-$ and $ZZ$), would be observable by IceCube. This leads to strong lower bounds on $\delta$ as a function of $m_\chi$. These bounds are shown in Fig.~\ref{fig:kinematic} for two choices of the loop-level spin-independent scattering cross section on nucleons, which is an additional free parameter that can alter the kinematics of the captured population and so affects the neutrino flux.

These bounds are relevant because if $\delta$ is too high, then it is possible for no incoming halo Higgsino to have sufficient energy to produce a nuclear recoil in the lab. We can therefore draw \textit{kinematic upper limits}, below which an inelastic DM particle is detectable via recoils on xenon (we choose the $^{136}$Xe isotope in this case since it has the largest kinematic reach). Scanning over all of our simulated halos, we obtain the band of possible upper limits shown in Fig.~\ref{fig:kinematic}. Looking at the $\sim 1$~TeV pure thermal Higgsino scenario, we see that only the extreme $\sim$2.5\% of halos would provide DM particles with enough energy to produce xenon recoils while satisfying the $\sigma_{\rm loop}^{\rm SI} = 0$ IceCube bound (that said, this model is also unlikely to provide a good fit to the data unless the recoil energy of the event also significantly underfluctuated within the expected energy resolution). See also Ref.~\cite{Ghosh:2026txe} for more discussion. On average, our halos yield a lower maximum speed on June 16, which is why we find that this kinematic constraint is slightly more severe for the simulations than it is for the SHM (comparing the solid and dashed lines).

\section{Conclusions}
\label{sec:conclusions}

If the 248~keV LZ nuclear-recoil candidate turns out to be the first sign of DM, it may have also brought with it a uniquely sensitive probe of the fastest particles in our galaxy's DM halo. Using 98 MW analogues from TNG-50, together with three higher-resolution FIRE-2 systems as a cross-check, we have examined the halo-to-halo variation of this extreme high-velocity tail to assess how much flexibility there is in the maximum possible DM speed.

Our main findings are as follows. First, the bulk local velocity distributions are very close to the SHM, consistent with previous analyses of TNG-50. On June 16, we find that the maximum laboratory-frame speed could plausibly have been between $697.1$ and $875.6$ km/s (95\% containment across simulations). Second, while the simulated DM halos generically exhibit pronounced co-rotation with the baryonic disk~\cite{Folsom:2026dqs}, we find that this does not translate into an equivalent reduction of the largest laboratory-frame speeds, because the local $v_\phi$ distribution is typically skewed rather than shifted by the co-rotation (see Figs.~\ref{fig:fvi} and~\ref{fig:summary_stats}).

These effects are directly relevant to endothermic inelastic DM interpretations of the LZ event. For TeV-scale DM and splittings of a few hundred keV, the recoil spectrum near $248~\mathrm{keV}$ is controlled by a sparsely populated and astrophysically uncertain portion of phase space. The typical parameter values required to explain the LZ event, e.g.~$m_\chi \sim 1$~TeV, $\delta \sim 375$~keV, put this model close to the kinematic boundary set by the maximum possible DM speed on June 16. The simulations we use have a shorter high-speed tail on average, and so this would somewhat alleviate the tension with the non-observation of higher-energy events in LZ.

If the LZ candidate is joined by more unexplained events with more exposure from current~\cite{XENON:2024wpa, PandaX:2018wtu} or next-generation experiments~\cite{Aalbers:2022dzr, XLZD:2024nsu}, then these issues should be testable through several additional observables. For instance, the shape of the recoil-energy spectrum and annual modulation are both sensitive to the high-velocity tail in different ways. The angular distribution of DM-induced recoils in this regime is also highly anisotropic, though searching for this class of signal will require more advanced detectors~\cite{Vahsen:2020pzb, OHare:2022jnx, Vahsen:2021gnb}.

\begin{acknowledgments}
I am supported by the Australian Research Council under grant numbers DE220100225 and CE200100008. I acknowledge the use of large language models in code development. All scientific judgments and any errors are my own.

\end{acknowledgments}

\bibliographystyle{bibi.bst}
\bibliography{bib.bib}

\end{document}